\documentclass[letterpaper, 10 pt, conference]{ieeeconf}
\usepackage{booktabs}
\IEEEoverridecommandlockouts                        
\usepackage{amsmath, amssymb, graphicx, hyperref}
\usepackage[ruled,vlined]{algorithm2e}
\usepackage{xcolor}
\usepackage{tikz}
\usetikzlibrary{shapes, arrows.meta, positioning, fit}

\newcommand\copyrighttext{%
\footnotesize \copyright 2026 IEEE. Personal use of this material is permitted. Permission from IEEE must be obtained for all other uses, in any current or future media, including reprinting/republishing this material for advertising or promotional purposes, creating new collective works, for resale or
redistribution to servers or lists, or reuse of any copyrighted component of this work in other works.}
\newcommand\copyrightnotice{%
\begin{tikzpicture}[remember picture,overlay]
\node[anchor=north,yshift=-20pt] at (current page.north) {\fbox{\parbox{\dimexpr\textwidth-\fboxsep-\fboxrule\relax}{\copyrighttext}}};
\end{tikzpicture}%
}
\usepackage{xcolor}

\usepackage{censor}
    
\title{\LARGE \bf
IMU-Centric Moving Horizon Estimation for Lateral Dynamics Estimation Across Vehicles and Grip Conditions}

\author{Seuffo Akouan'ha Ngoune$^{1}$, Alessandro Toschi$^{1}$, Paolo Burgio$^{1}$ and Marko Bertogna$^{1}$%
\thanks{$^{1}$University of Modena and Reggio Emilia, Italy,\newline
        {\tt\small \{seuffo.akouanhangoune, alessandro.toschi, paolo.burgio, marko.bertogna\}@unimore.it}}}

\begin{document}
\maketitle
\copyrightnotice
\thispagestyle{empty}
\pagestyle{empty}

\copyrightnotice


\begin{abstract}
Accurate estimation of lateral vehicle dynamics near the adhesion limit is important for stability control and high-performance driving, but lateral velocity is rarely measured directly because sensors such as optical sensors are costly. This paper presents an inertial measurement unit (IMU)-centric Moving Horizon Estimation framework that reconstructs lateral velocity using standard onboard signals, without relying on exteroceptive odometry or detailed tire-parameter tuning. Experimental validation on human-driven sports cars and an autonomous open-wheel race car across tracks, maneuvers, and conditions demonstrates accurate and robust lateral velocity and lateral acceleration estimates.
The proposed framework is available at \url{https://github.com/Aseuffo/IMU-Centric-MHE}      
\end{abstract}

\section{INTRODUCTION}
\label{section:intro}

Accurate estimation of lateral vehicle dynamics is essential for stability control and high-performance driving, especially near the tire friction limits, where force generation becomes strongly nonlinear. Longitudinal velocity is generally more accessible from signals already available on the vehicle, such as wheel-speed measurements and, when present, GNSS or other velocity estimates. Lateral velocity $v_y$, by contrast, is rarely measured directly in production vehicles due to the cost and practicality of dedicated sensors. As a result, $v_y$ is typically reconstructed from onboard measurements using model-based state estimation. Although adaptive schemes exist, model-based estimators typically remain sensitive to tire/road variability and to the assumptions embedded in the tire model. Changes in grip level, temperature, wear, or road surface can leave residual model mismatch, bias lateral-force predictions, and degrade $v_y$ estimates, especially near the friction limits where accurate lateral-state information is most needed. A common way to improve robustness is to complement inertial sensing with exteroceptive motion cues, such as visual or LiDAR odometry, and fuse them through observer-based estimators, including nonlinear Kalman-filter variants such as the EKF and UKF. This can yield strong performance, but it requires sensing modalities that are not always present on the vehicle and whose reliability may decrease under adverse environmental conditions. 

The main contribution of this paper is an \emph{IMU-centric} Moving Horizon Estimation (MHE) framework that reconstructs lateral velocity and lateral acceleration from inertial measurements and available onboard signals, while adapting tire force capacity to account for tire/road variability. The formulation also allows measured control inputs, such as steering and longitudinal acceleration, to be handled within the estimation window rather than being treated only as perfectly known exogenous quantities. The estimator combines a planar single-track model with a compact Sine Saturation Tire (SST) surrogate, where the effective front and rear force-capacity coefficients are estimated from a low-dimensional set of bounded and regularized decision variables. Building on our previous study based on a Pacejka-model UKF and odometry-aided updates (LOP-UKF)~\cite{lop_ukf}, we move from recursive filtering to windowed optimization, which makes it possible to impose physical bounds directly in the estimation problem. 

Section~\ref{sec:related_work} positions the proposed IMU-centric, bounded-parameter formulation with respect to observer, filtering, and external-odometry-aided pipelines. Section~\ref{section:vehicle_model} introduces the vehicle model and the SST tire surrogate, while Section~\ref{subsec:observability} discusses observability aspects of the IMU-centric setup. Section~\ref{section:mhe_formulation} presents the constrained MHE formulation. Section~\ref{section:results} reports experimental results and benchmarking on three vehicle platforms across two validation settings: the public \emph{REVS Program Vehicle Dynamics Database}, including a Ferrari 250 LM and a Corvette Grand Sport, and an autonomous Super Formula \emph{EAV-25}.

\begin{figure}[t] 
    \centering     
    \includegraphics[width=1.0\linewidth]{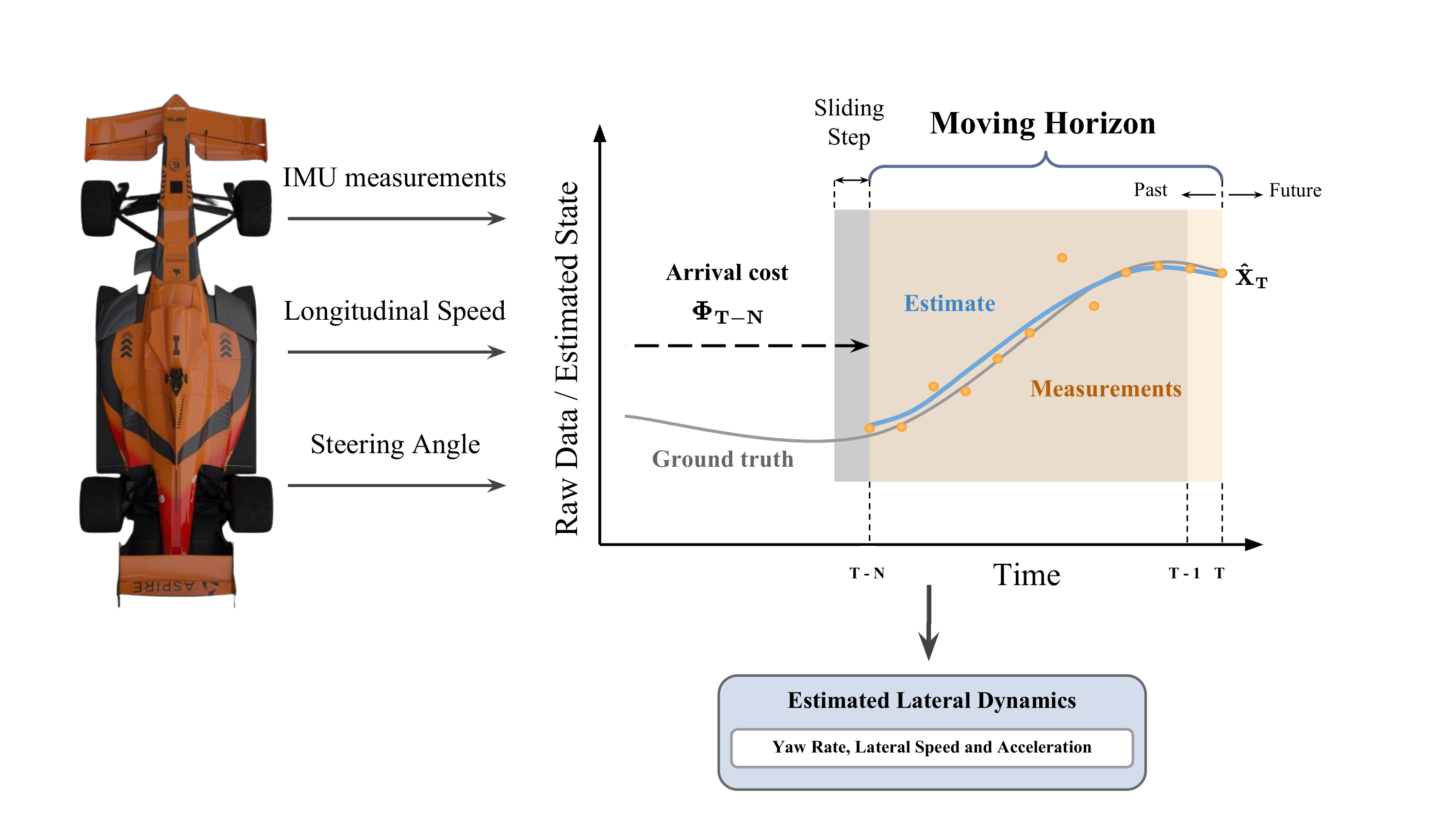}
    \caption{Moving Horizon Estimation scheme for lateral dynamics using onboard measurements. The sliding window with arrival cost provides yaw rate and lateral velocity estimates.} 
    \label{fig:a2rl_q1b} 
\end{figure}

\section{RELATED WORK}
\label{sec:related_work}

Estimating vehicle lateral velocity $v_y$ and sideslip angle $\beta$ has been extensively studied. Recent surveys cover model-based observers, Kalman-filter variants, and data-driven approaches~\cite{chindamo2018review}. A consistent takeaway is that performance depends strongly on excitation and on the adopted tire-force representation, especially when $v_y$ is not directly measured~\cite{grip2009}.

Some solutions use kinematic reconstruction and low-order observers because they are simple and inexpensive to run~\cite{selmanaj2017kinematic}. Their limitations become clear near the handling limits, where tire forces saturate and nonlinear effects dominate. Hybrid kinematic--dynamic formulations combine complementary model structures and tend to remain reliable across a broader set of maneuvers~\cite{villano2021crossukf}.

Among model-based estimators, recursive Bayesian filters remain widely used~\cite{chindamo2018review}. Extended Kalman Filter (EKF) designs are attractive for efficiency but can suffer from linearization errors in strongly nonlinear regimes. Unscented Kalman Filter (UKF) formulations reduce sensitivity to local linearization by propagating sigma points through the nonlinear dynamics, often improving robustness at the cost of higher computation~\cite{wan2000ukf,alshawi2024adaptiveukf}. To address parameter uncertainty and changing friction, adaptive UKF formulations update selected entries of the process and measurement noise covariances online, increasing measurement uncertainty in high-slip conditions and tuning the friction-related dynamics, modeled as a random-walk state, based on acceleration residual indicators, thereby improving robustness in extreme maneuvers~\cite{alshawi2024adaptiveukf}. A recurring challenge is that high-dimensional tire parameterizations, e.g., full Magic Formula identification, can be tightly coupled with slip reconstruction, leading to ill-conditioning when $v_y$ is unmeasured or weakly excited~\cite{grip2009,tripid2018}.

Several approaches improve observability by augmenting onboard sensing with exteroceptive sources. GNSS/INS fusion can provide global velocity cues~\cite{bevly2006insgps,ryu2002gps}, but GNSS signals can be sensitive to surroundings, e.g., heavy foliage and urban canyons, and may suffer temporary outages~\cite{leung2011gpsinsreview}. Vision-based methods can provide additional motion cues~\cite{serena2023vision,kuyt2018ccta}, while LiDAR-odometry-aided filtering has shown improved robustness under grip variations and imperfect tire tuning~\cite{lop_ukf}. Learning-based estimators, often recurrent networks, can achieve strong performance in specific regimes~\cite{srinivasan2020endtoend,giuliacci2023rnn} but may suffer from limited interpretability and reduced generalization outside the training distribution.

Finally, Moving Horizon Estimation (MHE) provides an optimization-based alternative by solving a constrained nonlinear estimation problem over a sliding window~\cite{rao2003nonlinearmhe,rawlings2017,diehl2009}. By explicitly enforcing physical constraints and bounds, MHE can be advantageous near saturation and under significant model mismatch~\cite{rawlings2017,diehl2009,canale2014dvsmhe}.

In contrast to pipelines that rely on external odometry sources or high-dimensional pre-calibrated tire models, the proposed approach addresses both limitations through an IMU-centric constrained MHE formulation with a compact SST surrogate and low-dimensional bounded force-capacity coefficients. The framework does not require prior lateral-velocity measurements for full tire-parameter fitting; instead, it relies on a small number of physically interpretable tuning choices, namely the SST curvature factor and admissible bounds on the force-capacity coefficients. This makes it suitable for vehicles and cost-limited experimental platforms where dedicated lateral-velocity sensing, detailed tire characterization, or additional odometry sources may be unavailable.

\section{SINE SATURATION TIRE MODEL}
\label{section:vehicle_model}

Moving Horizon Estimation (MHE) relies on the repeated solution of a nonlinear program; therefore, the process model must be differentiable and sufficiently lightweight for real-time implementation~\cite{rawlings2017,diehl2009}. We adopt a planar single-track model. Standard modeling assumptions and notation follow our previous works~\cite{lop_ukf,benchmark_mpc}.

\subsection{Planar single-track dynamics}
\label{subsec:single_track_ay_loads}

The vehicle is modeled as a rigid body moving in the horizontal plane with lumped front and rear axles on the centerline. Pitch, roll, and vertical dynamics are neglected, and the single-track yaw equation is used without an explicit differential yaw-moment term. Combined-slip effects, camber, and road banking are not modeled in the baseline formulation; however, banking can be incorporated through the load model or additional terms when the required information is available~\cite{benchmark_mpc}.

The state and input vectors are
\begin{equation}
x \triangleq \begin{bmatrix} v_x & v_y & r & a_y \end{bmatrix}^\top,
\qquad
u \triangleq \begin{bmatrix} \delta & a_x \end{bmatrix}^\top,
\end{equation}
where $v_x$ and $v_y$ are the longitudinal and lateral CoG velocities, $r$ is the yaw rate (rate of change of the vehicle heading $\psi$), $a_y$ is the lateral acceleration state, $\delta$ is the steering angle, and $a_x$ is the longitudinal acceleration input.

The continuous-time equations of motion are
\begin{equation}
\label{eq:bicycle_model}
\begin{aligned}
\dot{v}_x &= a_x + v_y r, \\
\dot{v}_y &= -v_x r + \frac{F_{y,f}\cos\delta + F_{y,r}}{m}, \\
\dot{r}   &= \frac{l_f F_{y,f}\cos\delta - l_r F_{y,r}}{I_z},
\end{aligned}
\end{equation}
with mass $m$, yaw inertia $I_z$, and CoG distances to the axles $l_f,l_r$. The lateral tire forces $F_{y,f}$ and $F_{y,r}$ are provided by the tire-force model described in Section~\ref{subsec:tire_model}.

To directly exploit the IMU lateral acceleration measurement in the MHE cost, $a_y$ is included as a state and defined by the lateral force balance
\begin{equation}
\label{eq:ay_def}
a_y(x,u,\theta) \triangleq \frac{F_{y,f}(x,u,\theta)\cos\delta + F_{y,r}(x,u,\theta)}{m},
\end{equation}
where $\theta=[\eta_{y,f},\eta_{y,r}]^\top$ collects the tire-model parameters.
Under a zero-order-hold (piecewise-constant input) assumption over one sampling interval~\cite{rawlings2017},
the dynamics of $a_y$ are approximated via the chain rule.
Let $\xi \triangleq [v_x\ v_y\ r]^\top$, then
\begin{equation}
\label{eq:ay_dyn}
\dot a_y \approx \nabla_{\xi} a_y(x,u,\theta)^\top\, \dot{\xi}
= \frac{\partial a_y}{\partial v_x}\dot v_x + \frac{\partial a_y}{\partial v_y}\dot v_y + \frac{\partial a_y}{\partial r}\dot r,
\end{equation}
This yields a differentiable expression consistent with the force-based dynamics in  \eqref{eq:bicycle_model} while allowing \(a_y\) to be directly constrained and regularized within the estimator.

Axle normal loads $F_{z,f}$ and $F_{z,r}$ enter the tire-force model through the peak-force scaling. In this work, they are computed using the same approximation adopted in~\cite{lop_ukf,benchmark_mpc}, namely a static load distribution with optional contributions from aerodynamics and longitudinal load transfer. When vehicle-specific parameters are unavailable (e.g., aerodynamics-related coefficients, load-transfer parameters such as center of gravity (CoG) height, or banking information), the load model is simplified accordingly; in the most reduced form only the static distribution is retained,
\begin{equation}
\label{eq:static_loads}
F_{z,f} = \frac{l_r}{l_f+l_r}\,mg,
\qquad
F_{z,r} = \frac{l_f}{l_f+l_r}\,mg,
\end{equation}
where $g$ is the gravitational acceleration, and aerodynamic, banking-related, and load-transfer terms are neglected.

\subsection{Simplified lateral tire-force model}
\label{subsec:tire_model}

Although high-fidelity tire models are available, jointly estimating a high-dimensional Magic-Formula parameter set together with slip angles becomes strongly coupled when $v_y$ is not directly measured, and can be ill-conditioned under low excitation.

To keep the MHE optimization well-conditioned, we adopt a compact differentiable surrogate with bounded force output, referred to here as the SST model. Slip angles are computed using the standard single-track congruence \begin{equation}
\label{eq:slip_angles}
\begin{aligned}
\alpha_f &= \delta - \arctan\!\left(\frac{v_y + l_f r}{v_x}\right),\\
\alpha_r &= -\,\arctan\!\left(\frac{v_y - l_r r}{v_x}\right).
\end{aligned}
\end{equation}
In implementation, $v_x$ is clamped to avoid numerical issues at very low speed.

For each axle $i\in\{f,r\}$, the SST lateral force model is
\begin{equation}
\label{eq:sst_model}
F_{y,i} = \eta_{y,i}\,F_{z,i}\,\sin(B_i\alpha_i),
\qquad i\in\{f,r\},
\end{equation}
where $B_i$ is a fixed curvature factor from nominal characterization, and $\eta_{y,i}>0$ is an effective lateral force-capacity coefficient estimated online.
Intuitively, $\eta_{y,i}$ scales the achievable force envelope (e.g., $|F_{y,i}|\le \eta_{y,i}F_{z,i}$), while also absorbing residual modeling errors and variability not represented by the surrogate. For this reason, $\eta_{y,i}$ belongs to the SST surrogate and should not be interpreted as a Magic-Formula parameter. In the MHE, $\eta_{y,i}$ is treated as a bounded and regularized decision variable to prevent drift under low excitation.

\begin{figure}[htb!]
    \centering
    \includegraphics[width=1.0\linewidth]{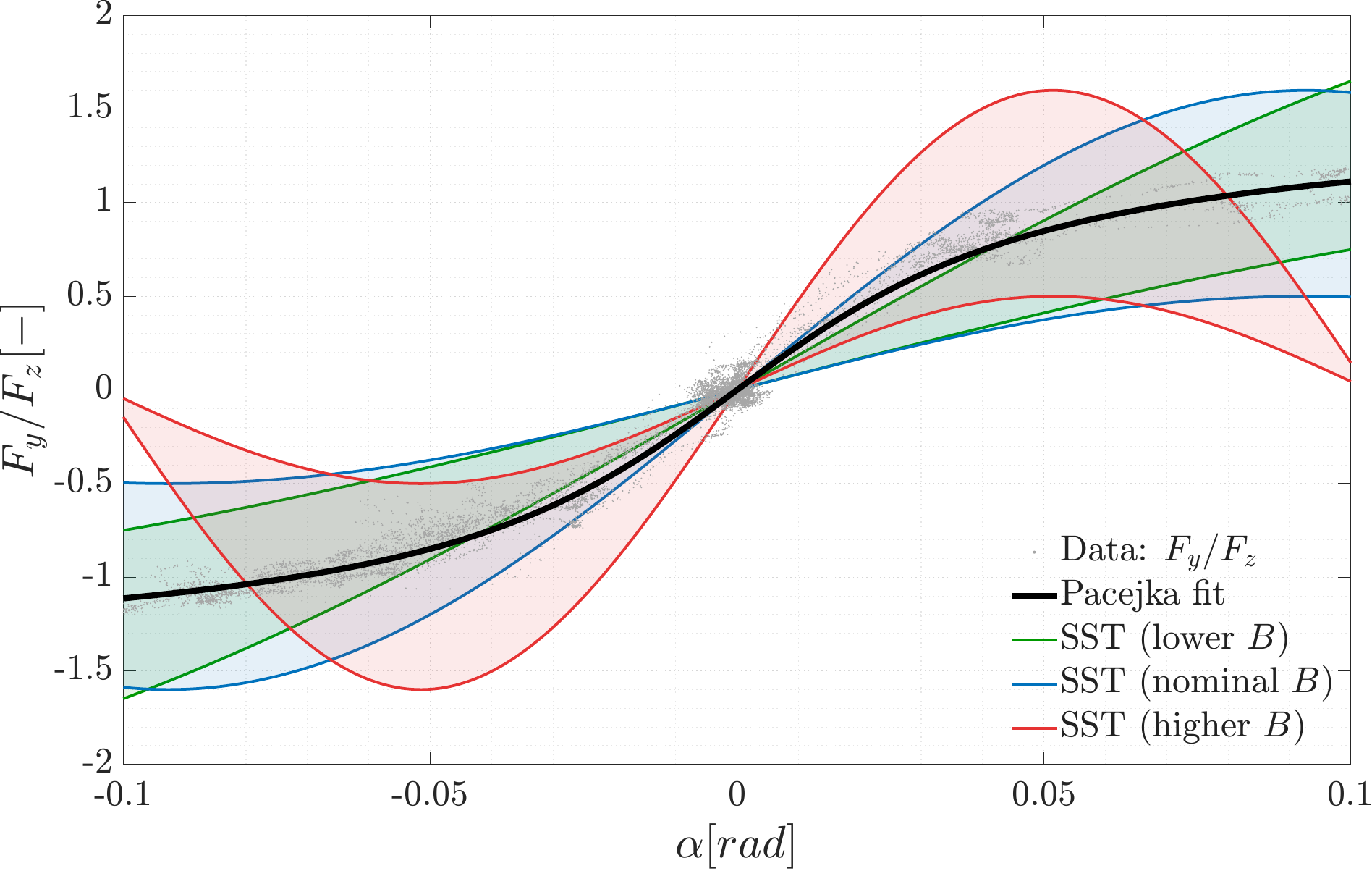}
    \caption{Normalized lateral force versus slip angle. The black curve is the Pacejka Magic Formula fit; the SST feasible envelope is generated by varying $\eta$ between the optimization bounds, for three cases of B.}
    \label{fig:curves}
\end{figure}

The SST model has the following properties that are useful for optimization-based estimation:
\begin{itemize}
    \item[(i)] \emph{Small-slip linearization.} For $|B_i\alpha_i|\ll 1$, $\sin(B_i\alpha_i)\approx B_i\alpha_i$, hence
    \begin{equation}
    \label{eq:sst_linearization}
    F_{y,i} \approx \eta_{y,i}\,B_i\,F_{z,i}\,\alpha_i,
    \end{equation}
    and the corresponding small-slip cornering stiffness is $C_{\alpha,i}=\eta_{y,i}B_iF_{z,i}$.
    
    \item[(ii)] \emph{Smooth bounded magnitude.} Since $|\sin(\cdot)|\le 1$ and $F_{z,i}\ge 0$, the force magnitude satisfies
    \begin{equation}
    \label{eq:sst_bound}
    |F_{y,i}| \le \eta_{y,i}\,F_{z,i},
    \end{equation}
    where the right-hand side follows from the enforced constraint $\eta_{y,i}>0$. Imposing an upper bound $\eta_{y,i}\le \bar{\eta}_i$ further yields a global cap $|F_{y,i}|\le \bar{\eta}_i F_{z,i}$, providing a smooth force envelope without additional shape/shift parameters.

    \item[(iii)] \emph{Validity range.} The mapping $\sin(B_i\alpha_i)$ is monotone for $|B_i\alpha_i|<\pi/2$ and becomes
    non-monotone beyond this interval due to periodicity. In this work, $B_i$ is selected such that $\pi/(2B_i)$ lies
    beyond the slip-angle envelope observed in the experimental datasets.
    
    \item[(iv)] \emph{Scalar amplitude correction and excitation.} With fixed $B_i$, the coefficient $\eta_{y,i}$ enters
    \eqref{eq:sst_model} as a multiplicative scale factor, thus acting as an amplitude correction over the slip-angle distribution observed within the estimation window. Moreover, $\partial F_{y,i}/\partial \eta_{y,i}=F_{z,i}\sin(B_i\alpha_i)$ becomes small when slip excitation is weak (small $|\alpha_i|$), motivating boundedness and regularization of $\eta_{y,i}$ to prevent drift under low excitation.
\end{itemize}

These considerations are illustrated in Fig.~\ref{fig:curves}, which shows the SST feasible envelopes obtained by sweeping the amplitude parameter $\eta$ within the bounds in~\eqref{eq:mhe_param_bounds} and by perturbing $B$ by $\pm 50\%$ around its nominal value, while remaining within the model validity range.

\section{OBSERVABILITY CONSIDERATIONS}
\label{subsec:observability}

In a nonlinear setting, \emph{observability} formalizes whether the internal state can be uniquely determined from measured outputs over a finite time interval under known inputs. Formally, a system is locally observable at $x_0$ if every neighboring initial state produces a distinguishable output trajectory for some admissible input~\cite{hermann1977}.
Since MHE reconstructs unmeasured components by fitting a dynamic model to measured trajectories over a window, its practical performance is inherently linked to local observability, and (when unknown parameters are included) to local identifiability.

\subsection{Nonlinear local observability and OI rank conditions}
For analytic nonlinear systems, local observability can be assessed via rank conditions based on Lie derivatives of the output map~\cite{hermann1977,villaverde2016}.
Given  the dynamic model $\dot{x}=f(x,u,\theta)$ and the measurement model $y=h(x,u)$, the Lie derivative of $h$ along $f$ is defined as
\begin{equation}
L_f h(x) \triangleq \frac{\partial h(x,u)}{\partial x}\, f(x,u,\theta),
\end{equation}
and recursively $L_f^{(i)}h(x)$. Stacking Jacobians of $h, L_f h, L_f^{(2)}h,\dots$ yields the nonlinear observability matrix.
The classical Observability Rank Condition (ORC) states that if this matrix has full column rank at $x_0$, the system is locally observable around $x_0$ ~\cite{hermann1977}.

To include constant unknown parameters, one augments the state with $\theta$ and assigns zero dynamics $\dot{\theta}=0$.
The corresponding generalized observability--identifiability (OI) matrix is constructed analogously, and the Observability--Identifiability Condition (OIC) states that local state observability and parameter identifiability hold when the OI matrix has full column rank $n+q$ at the operating point~\cite{villaverde2016}.

\subsection{Measurement set, unknowns, and derivative order}
Our measured outputs are $y=[v_x,\;r,\;a_y]^\top$ with known inputs $u=[\delta,\;a_x]^\top$.
The quantity of interest is the lateral velocity $v_y$, while the effective SST force-capacity coefficients
$\theta=[\eta_{y,f},\eta_{y,r}]^\top$ are treated as unknown constant parameters.
Accordingly, we consider the augmented vector
\begin{equation}
\tilde x = [\,v_x,\; v_y,\; r,\; \eta_{y,f},\; \eta_{y,r}\,]^\top,
\end{equation}
with $(n+q)=5$ unknowns and $m=3$ outputs. As noted in the OIC construction, each Lie-derivative block contributes $m$ rows; therefore, the minimum Lie-derivative order for which the OI matrix may \emph{possibly} become full column rank is~~\cite{villaverde2016}
\begin{equation}
\label{eq:nd_min}
n_d=\left\lceil\frac{n+q}{m}\right\rceil-1
=\left\lceil\frac{5}{3}\right\rceil-1=1.
\end{equation}
Moreover, Lie derivatives of order higher than $(n+q-1)$ cannot increase the rank~~\cite{villaverde2016}.
This provides a bounded procedure: add Lie-derivative blocks until either full rank is reached or the rank ceases to increase.

\subsection{Local observability of lateral velocity}
Although $a_y$ is implemented as an augmented state for numerical purposes, it is constrained to remain consistent with the force-based expression \eqref{eq:ay_def}; hence, for observability analysis we consider the reduced representation $a_y=a_y(x,u,\theta)$.
Since $v_x$ and $r$ are directly measured, the key question is whether the measured lateral acceleration $a_y$ carries local information about $v_y$ through the slip angles in \eqref{eq:slip_angles}.
A sufficient local condition is
\begin{equation}
\frac{\partial a_y}{\partial v_y}\neq 0.
\label{eq:ay_vy_sensitivity}
\end{equation}
This condition holds generically whenever $v_x>0$ and the lateral force map is locally monotone in the operating region (Section~\ref{subsec:tire_model}): both slip angles $\alpha_f,\alpha_r$ depend on $v_y$ and enter $a_y$ through the axle lateral forces, so the chain-rule sensitivity is nonzero except at very low speed or at flat-slope points of the surrogate force map.

\subsection{Local identifiability of tire parameters}
Local identifiability of $\theta$ additionally requires that front and rear force capacities affect the measured signals in linearly independent ways.
Using the SST law $F_{y,i}=\eta_{y,i}F_{z,i}\sin(B_i\alpha_i)$ with $i\in\{f,r\}$, the sensitivities of lateral acceleration and yaw acceleration with respect to $\eta_{y,i}$ can be written compactly as
\begin{equation}
\label{eq:theta_sens_compact}
\begin{aligned}
\frac{\partial a_y}{\partial \eta_{y,i}} &= \frac{1}{m}\,\gamma_i\,F_{z,i}\sin(B_i\alpha_i),\\
\frac{\partial \dot r}{\partial \eta_{y,i}} &= \frac{1}{I_z}\,\rho_i\,F_{z,i}\sin(B_i\alpha_i),
\qquad i\in\{f,r\}.
\end{aligned}
\end{equation}
where $\gamma_f=\cos\delta$, $\gamma_r=1$, $\rho_f=l_f\cos\delta$, and $\rho_r=-l_r$.
Hence, $\eta_{y,f}$ and $\eta_{y,r}$ are locally identifiable when both axles experience non-negligible slip excitation (i.e., $\sin(B_f\alpha_f)\neq 0$ and $\sin(B_r\alpha_r)\neq 0$) and $\cos\delta\neq 0$.
Intuitively, lateral acceleration provides a force-sum constraint, while yaw dynamics provide a force-moment constraint, allowing separation of front and rear contributions when the vehicle is sufficiently excited.

\subsection{Degenerate regimes and mitigation in MHE}
Loss of observability and identifiability is expected under weak lateral excitation, notably:
(i) very low speed ($v_x\approx 0$) and/or (ii) near-straight driving with small steering and negligible slip ($\alpha_f\approx\alpha_r\approx 0$), for which $a_y\approx 0$ and the parameter sensitivities in \eqref{eq:theta_sens_compact} collapse.
These regimes are handled in the proposed MHE by bounding and regularizing $\eta_{y,i}$ (Section~\ref{section:mhe_formulation}), which prevents drift when the data are temporarily uninformative, while focusing evaluation on datasets containing sustained lateral maneuvers where the above conditions hold.

\section{MOVING HORIZON ESTIMATION}
\label{section:mhe_formulation}

Moving Horizon Estimation (MHE) reconstructs the current state by repeatedly solving a constrained optimization problem over a sliding time window~\cite{rao2003nonlinearmhe,rawlings2017,diehl2009}.
Compared to recursive filters, this formulation naturally supports state/parameter bounds and additional physical constraints, which is beneficial when nonlinearities and saturation effects dominate.

\subsection{MHE versus Kalman Filtering}
\label{subsec:mhe_vs_kf}

MHE and Kalman filtering are closely related: in the linear--Gaussian and unconstrained case, the MHE solution can be
made equivalent to the Kalman filter estimate through an appropriate arrival cost, so the practical question is not
``which estimator is better'' in general, but which formulation best matches the structure of the target problem~\cite{haseltine2005,rawlings2017}.

In our application, two aspects motivate MHE:
\begin{itemize}
\item \textit{Constraints:} bounds on states and on the SST parameters are physically meaningful and directly enforceable in MHE, whereas standard EKF/UKF treat them only heuristically~\cite{rawlings2017,diehl2009}.
\item \textit{Nonlinearity and saturation:} near-limit driving and bounded tire-force behavior reduce the validity of local linearization assumptions underlying EKF, while MHE can accommodate these regimes through repeated constrained nonlinear optimization~\cite{rawlings2017}.
\end{itemize}

\subsection{Discrete-Time Estimation Model}
\label{subsec:mhe_model}

The estimation model is the discrete-time counterpart of the continuous dynamics in
Section~\ref{section:vehicle_model} and is written as
\begin{align}
x_{k+1} &= f_d(x_k,\tilde u_k,\theta) + w_k, \label{eq:mhe_discrete_process}\\
y_k     &= h(x_k) + v_k, \label{eq:mhe_measurement}
\end{align}
where $x_k$ is the state, $\tilde u_k$ is the input used by the dynamic model, $y_k$ is the measurement, $\theta$ is a possibly unknown parameter vector, and $w_k$, $v_k$ denote process and measurement disturbances.

The measured input vector is denoted by
\begin{equation}
u_k^m =
\begin{bmatrix}
\delta_k^m & a_{x,k}^m
\end{bmatrix}^{\top},
\end{equation}
where $\delta_k^m$ and $a_{x,k}^m$ are the measured steering angle and longitudinal acceleration. Since these signals may contain delay, calibration errors, or measurement noise, the estimator introduces a small input correction
\begin{equation}
\tilde u_k = u_k^m + \nu_k,
\end{equation}
where $\nu_k$ is treated as an optimization variable and penalized in the MHE cost. The vehicle dynamics are therefore evaluated using $\tilde u_k$ rather than the raw measured input $u_k^m$.

The transition map $f_d(\cdot)$ is obtained by numerically integrating the continuous-time model over one sampling
interval $T_s$ under a zero-order hold on $\tilde{u}_k$; throughout, we use a fourth-order Runge--Kutta discretization (RK4)~~\cite{butcher2016}. In the MHE implementation, we adopt a multiple-shooting transcription~\cite{diehl2009} to improve numerical robustness and enable efficient warm-starting in receding-horizon operation.

\subsection{MAP Viewpoint and MHE Cost}
\label{subsec:mhe_map}

Consider a sliding window of length $N$ ending at time $k$. Moving Horizon Estimation (MHE) can be interpreted as a
maximum \emph{a posteriori} (MAP) estimate of the state trajectory $x_{k-N:k}$, the input-correction sequence
$\nu_{k-N:k-1}$, and a constant parameter vector $\theta$ over the window~\cite{rao2003nonlinearmhe,rawlings2017}.
Under approximately Gaussian disturbances, the negative log-posterior leads to the weighted least-squares problem
\begin{equation}
\label{eq:mhe_cost_generic}
\begin{aligned}
\min_{\substack{x_{k-N:k}\\ \theta\\ \varepsilon_{k-N:k-1}\\ \nu_{k-N:k-1}}}\;\;
&\|x_{k-N}-\hat{x}_{k-N}\|_{P}^2
+\sum_{i=k-N}^{k-1}
\bigl(\|\varepsilon_i\|_{Q}^2+\|\nu_i\|_{U}^2\bigr)\\
&+\sum_{i=k-N}^{k}\|y_i-h(x_i)\|_{R}^2
+\|\theta-\hat{\theta}\|_{S}^2 .
\end{aligned}
\end{equation}
subject to the dynamic and physical constraints
\begin{align}
x_{i+1} &= f_d(x_i,\tilde u_i,\theta) + \varepsilon_i,
&& i=k-N,\dots,k-1, \label{eq:mhe_dyn_slack}\\
u_{\min} &\le \tilde u_i \le u_{\max},
&& i=k-N,\dots,k-1, \label{eq:mhe_input_bounds}\\
x_{\min} &\le x_i \le x_{\max},
&& i=k-N,\dots,k, \label{eq:mhe_state_bounds}\\
\theta_{\min} &\le \theta \le \theta_{\max}.
&& \label{eq:mhe_param_bounds}
\end{align}
Here $\varepsilon_i$ is the model-consistency residual that relaxes the dynamics to account for disturbances, while $\nu_i$ is the input-correction residual applied to the measured input $u_i^m$. The notation $\|e\|_W^2 := e^\top W e$ is used, and $P,Q,R,S,U \succ 0$ are positive-definite weighting matrices.
The matrix $P$ weights the arrival cost, $R$ penalizes measurement residuals, $Q$ penalizes the model-consistency residuals $\varepsilon_i$, $S$ regularizes the parameter deviation from its prior $\hat{\theta}$,
and $U$ penalizes the input corrections $\nu_i$.
Thus, larger values of $U$ force the corrected input $\tilde u_i$ to remain close to the measured input $u_i^m$, whereas smaller values
allow greater compensation for input uncertainty
~\cite{rao2003nonlinearmhe,rawlings2017}.

\subsection{Implementation, Solver, and Warm Start}
\label{subsec:mhe_impl}

The MHE nonlinear program is implemented in \texttt{CasADi}, which provides automatic differentiation and NLP construction~\cite{casadi2019}.
The resulting problem is solved using \texttt{IPOPT}, a primal--dual interior-point optimizer~\cite{wachter2006ipopt}.
Warm-start settings (e.g., \texttt{warm\_start\_init\_point}) follow the \texttt{IPOPT} documentation~\cite{ipopt_warmstart}.

Since consecutive MHE problems differ only by a one-step window shift, we apply warm starting by:
(i) shifting the previous primal solution (state and input trajectories) to initialize the next solve; and
(ii) reusing the previous Lagrange multipliers (denoted generically by $\lambda$) associated with constraints and bounds.
This reduces the number of iterations and improves real-time feasibility in receding-horizon operation.

\section{RESULTS}
\label{section:results}

We present a set of results validating the proposed MHE approach using data from three different vehicles across two
racing tracks. The following subsections highlight key strengths of the method, emphasizing stability under varying
modeling assumptions and operating conditions.

All results are reported offline; however, measured solver runtimes suggest real-time feasibility on a high-end CPU.
With $T_s=0.01\,\mathrm{s}$, the average solve time was $3.9\,\mathrm{ms}$ for $N=10$ (range $3.65$--$4.68\,\mathrm{ms}$)
and $6.3\,\mathrm{ms}$ for $N=30$ (range $5.98$--$8.25\,\mathrm{ms}$) on an Intel\textsuperscript{\textregistered}
Core\texttrademark{} i9-14900HX, leaving margin within the sampling period.
\begin{figure}[htb!]
    \centering
    \includegraphics[width=1.0\linewidth]{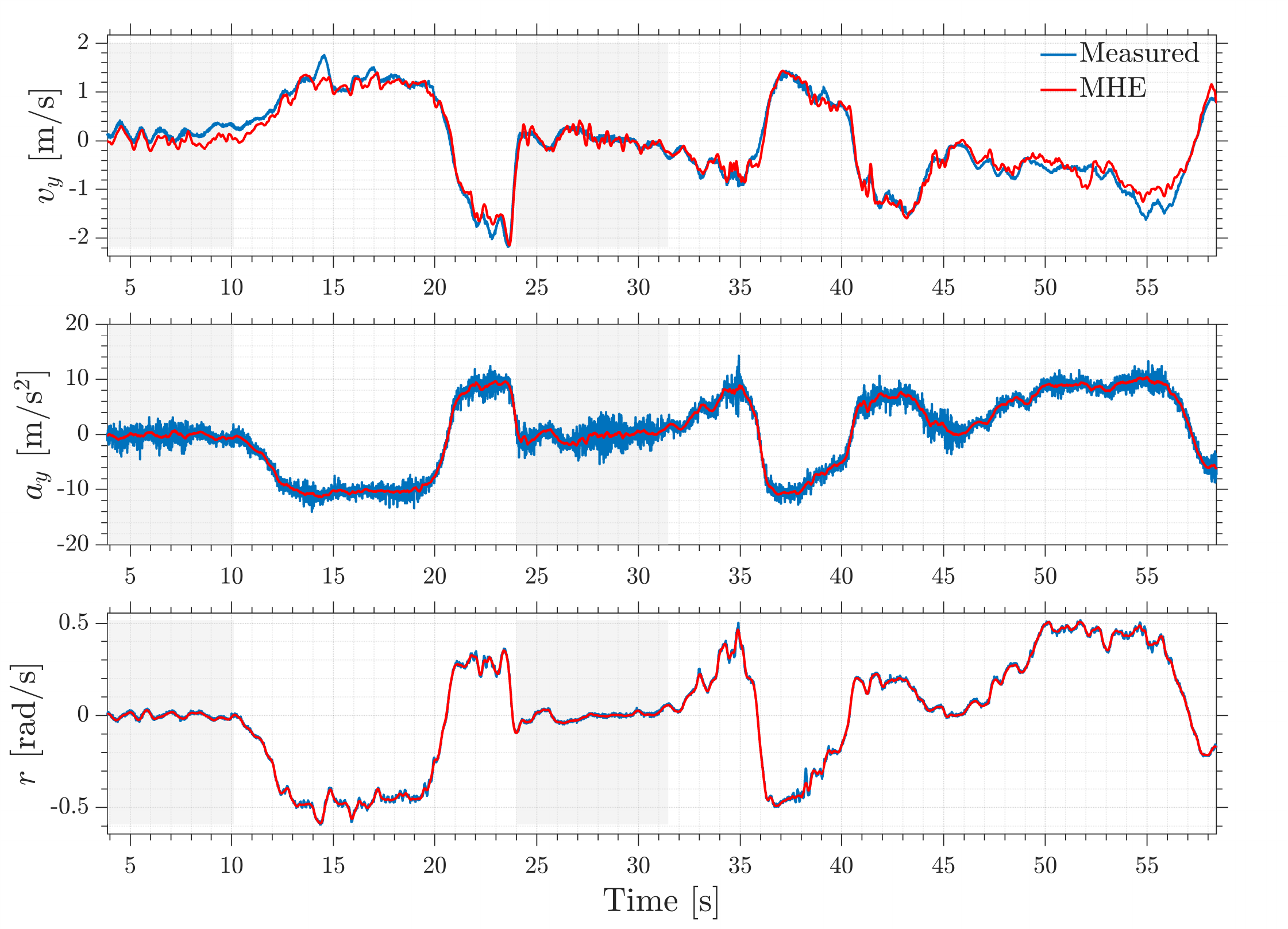}
    \caption{Estimation of lateral dynamics during the 22 Feb 2014 session (run ID: 02\_01\_03), laps 5--6. The MHE estimate (red) closely tracks the ground truth (blue).}
    \label{fig:a2rl_q1}
\end{figure}
The first two cases analyze open-source data\footnote{https://purl.stanford.edu}
from the 1965 Ferrari 250 LM and the 1963 Corvette Grand Sport, both driven at Palm Beach International Raceway. In these
experiments, aerodynamic effects and longitudinal load transfer were neglected, as the corresponding parameters were not available in the dataset. As discussed in the model description section, these terms are not strictly required for the MHE framework to operate effectively. The yaw moment of inertia was also not provided and was therefore approximately estimated using the geometric dimensions and total mass of the vehicles. Although several missing parameters could have been identified through data-driven model fitting, we deliberately limited the model to geometric dimensions and mass
only. Steering and heading offsets were also neglected. This choice was made to demonstrate the ability of the framework to generalize and provide reliable estimates even in the presence of modeling inaccuracies.

The last case instead presents results obtained using data from the autonomous Super Formula EAV-25 during a qualifying session of the Abu Dhabi Autonomous Racing League\footnote{https://a2rl.io/}, conducted on the North Layout of the Yas Marina Circuit. In this scenario, the complete vehicle model was employed, using accurately identified parameters validated by the team through extensive experimental testing.
In the REVS cases, the reference lateral velocity was obtained from the post-processed dual-antenna GNSS/INS data provided with the dataset, while the Super Formula EAV-25 reference $v_y$ was provided by an optical velocity sensor.

\subsection{Case 1: Ferrari state estimation}

The first analyzed case corresponds to a nominal racing scenario in which the MHE was initialized with effective SST lateral force-capacity coefficients $\eta_{y,i}$ close to their nominal operating values. The moving horizon window length was set to $0.1\,\text{s}$, corresponding to 10 discretization steps.

As shown in Fig.~\ref{fig:a2rl_q1}, the MHE accurately estimates the lateral velocity $v_y$, matching the post-processed ground truth even during aggressive cornering. The estimated lateral acceleration $a_y$ and yaw rate $r$ follow the measured signals while effectively filtering out IMU noise.
The dark grey shaded regions highlight portions of the run where $ |a_y| < 1\,\text{m/s}^2$, corresponding to low-excitation conditions. These segments are included to demonstrate the stability of the estimator even when the vehicle operates at very small slip angles, as discussed in the observability section \ref{subsec:observability}.

\subsection{Case 2: Corvette Tire Analysis}
\begin{figure}[htb!]
    \centering
    \includegraphics[width=1.0\linewidth]{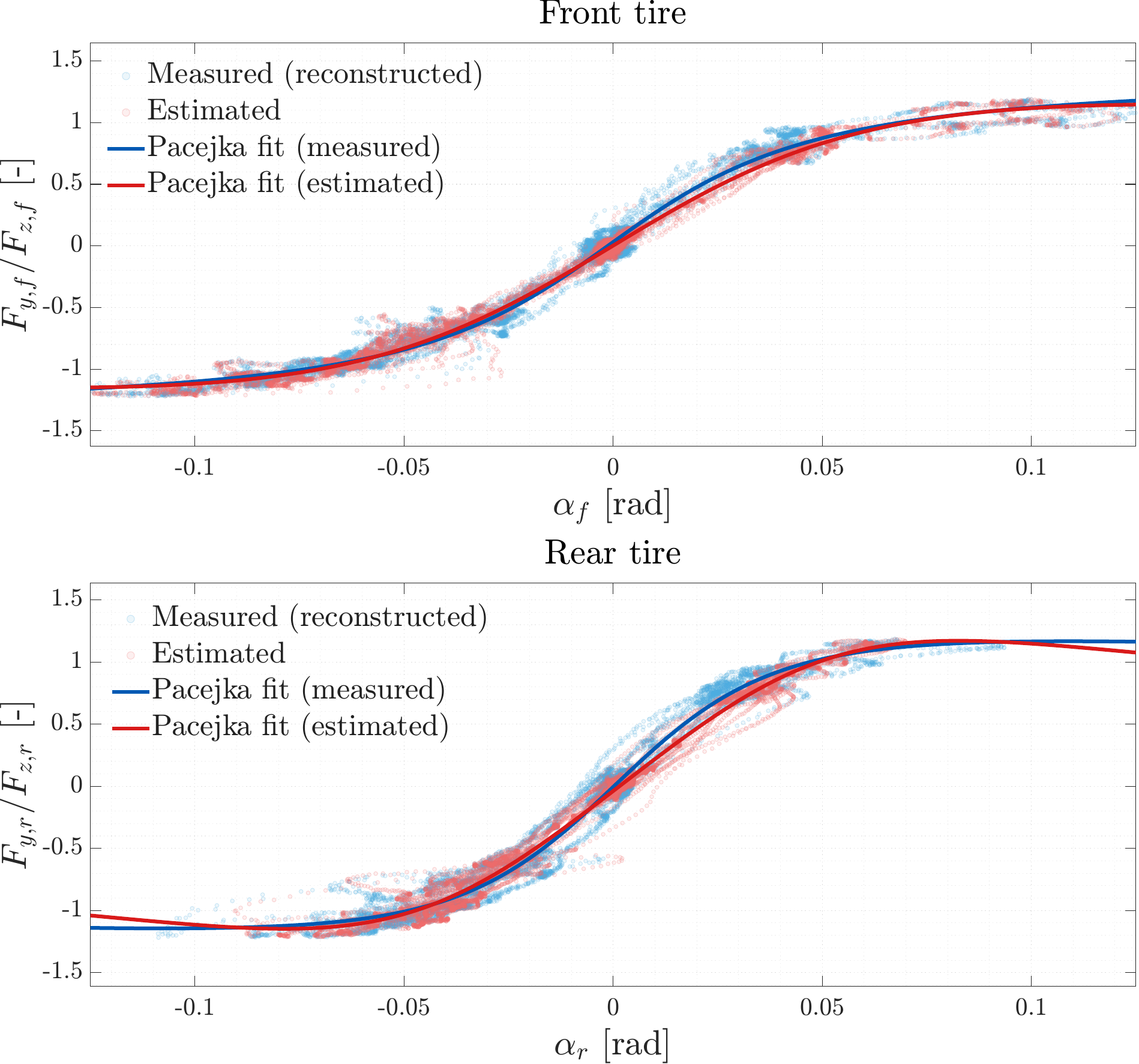}
    \caption{Comparison between reconstructed and estimated tire lateral forces for front and rear  axles. The MHE accurately tracks the nonlinear tire behavior, with the Pacejka fits showing high consistency between measured data  and estimator outputs.}
    \label{fig:pacejka}
\end{figure}

In this scenario, we investigate how lateral dynamics estimation can be exploited to retrieve tire characteristics. The Corvette dataset (23 Feb 2013 session; run ID: \texttt{01\_01\_03}; laps~2--4) was analyzed by reconstructing the normalized lateral tire forces as a function of slip angle for both the front and rear axles (Fig.~\ref{fig:pacejka}).

The blue scatter represents the tire forces computed using filtered IMU measurements together with the ground-truth lateral velocity. The red scatter instead corresponds to the reconstruction obtained using the MHE-estimated lateral acceleration $a_y$ and lateral velocity $v_y$.

The MHE-based reconstruction accurately captures the tire behavior across both the linear and saturation regimes. Remarkably, even though the estimator relies on a simplified vehicle model, the nonlinear tire characteristics naturally emerge in the reconstructed forces. This highlights the robustness of the MHE framework in preserving physically consistent nonlinear behavior despite modeling approximations.
To further quantify this agreement, two Pacejka curves were independently fitted to the two scatters using nonlinear least-squares regression. The resulting fits show strong overlap, confirming that the MHE provides sufficiently accurate state estimates to enable reliable tire parameter identification.

\subsection{Case 3: Autonomous Super Formula Analysis}

A key robustness test consists of initializing the estimator with incorrect tire-related parameters, in particular the SST effective force-capacity coefficient $\eta_{y,i}$. This coefficient captures the available lateral force envelope under the current operating conditions (e.g., grip, and other unmodeled effects) and therefore cannot be assumed known a priori. Within a few iterations, the MHE adapts $\eta_{y,i}$ to match the measured vehicle response.
\begin{figure}[htb!]
    \centering
    \includegraphics[width=1.0\linewidth]{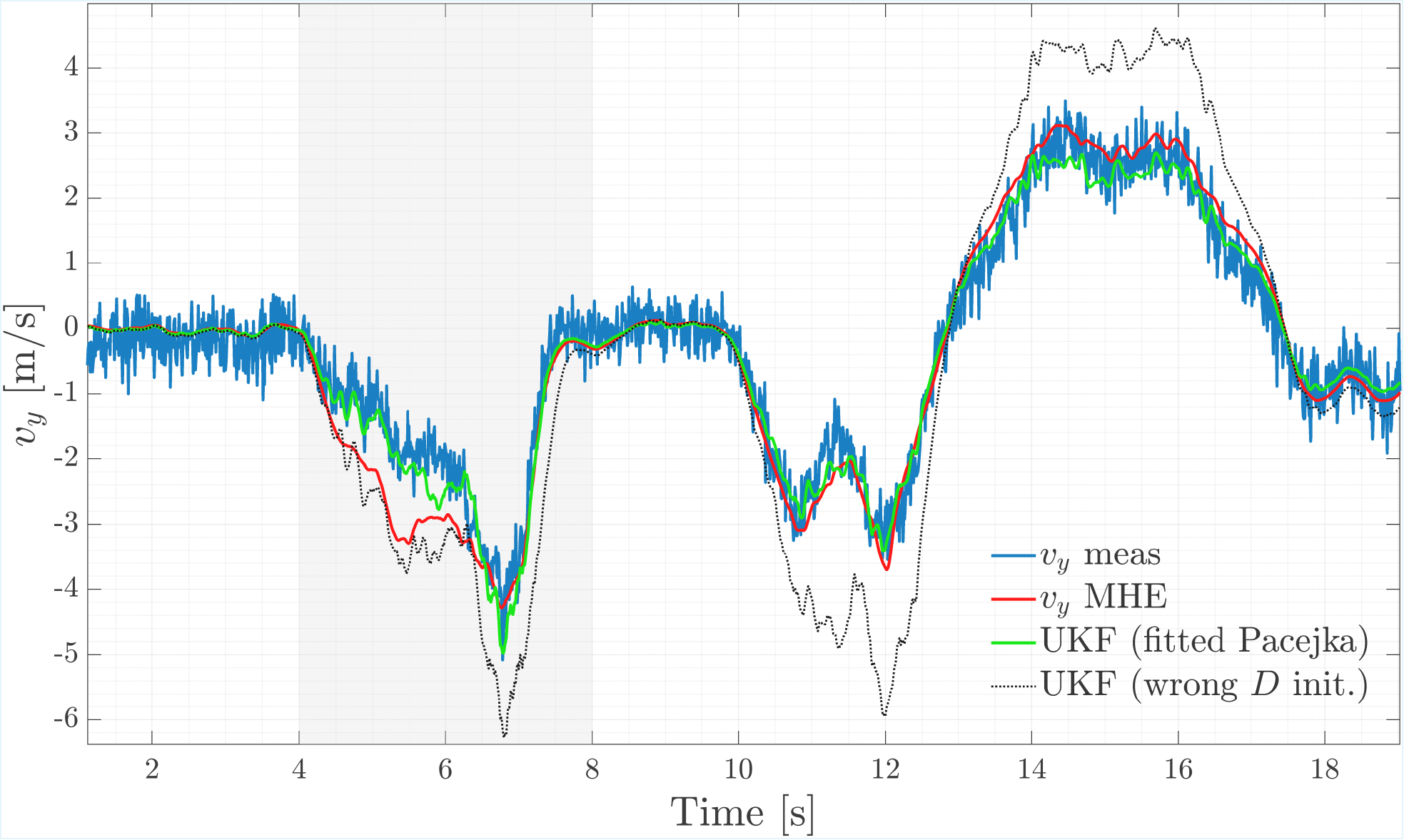}
    \caption{Convergence analysis on the first three corners of a Super Formula qualifying lap (Q1). The MHE (red), initialized with underestimated $\eta_{y,i}$, rapidly converges to the correct state trajectory. It is compared with a UKF using a posteriori fitted tire parameters (green) and a UKF initialized with reduced Pacejka peak factor $D$ (dashed black).}
    \label{fig:convergence}
\end{figure}

The adaptation rate is mainly influenced by (i) the distance between the initial guess and the effective value during the run, and (ii) the arrival-cost weighting, which regulates how strongly past estimates constrain the current window.

A larger arrival-cost weight increases confidence in past information and typically slows down parameter adaptation.
Conversely, a smaller arrival-cost weight accelerates convergence by emphasizing recent measurements, but if set too low it may amplify measurement noise and reduce long-term consistency.

Figure~\ref{fig:convergence} shows the first three corners of the Super Formula EAV-25 during a qualifying lap (Q1), where grip conditions were significantly lower than in subsequent sessions. The coefficient $\eta_{y,i}$ was deliberately initialized to a value substantially below its effective level. As shown in Fig.~\ref{fig:convergence}, the MHE rapidly compensates for this mismatch through the optimization and converges to the correct lateral velocity $v_y$ within a fraction of a second. The shaded grey region highlights Turn~1: although the response is initially overestimated at corner entry, convergence is already achieved by the apex.

Two additional curves correspond to a UKF relying purely on the model (LOP-UKF without exteroceptive updates). In the green curve, tire parameters were identified a posteriori using data from the same lap and the filter achieves excellent
accuracy. In the dashed black curve, the UKF was initialized with an underestimated Pacejka peak factor ($D$), which induces a persistent bias that cannot be compensated online.

While the UKF with correctly identified tire parameters achieves excellent performance, the MHE attains comparable accuracy after only one corner, without prior knowledge of the exact friction level. This behavior enables indirect, online estimation of the track grip conditions through the adaptation of the SST force-capacity coefficient $\eta_{y,i}$.

\begin{figure}[htb!]
    \centering
    \includegraphics[width=1.0\linewidth]{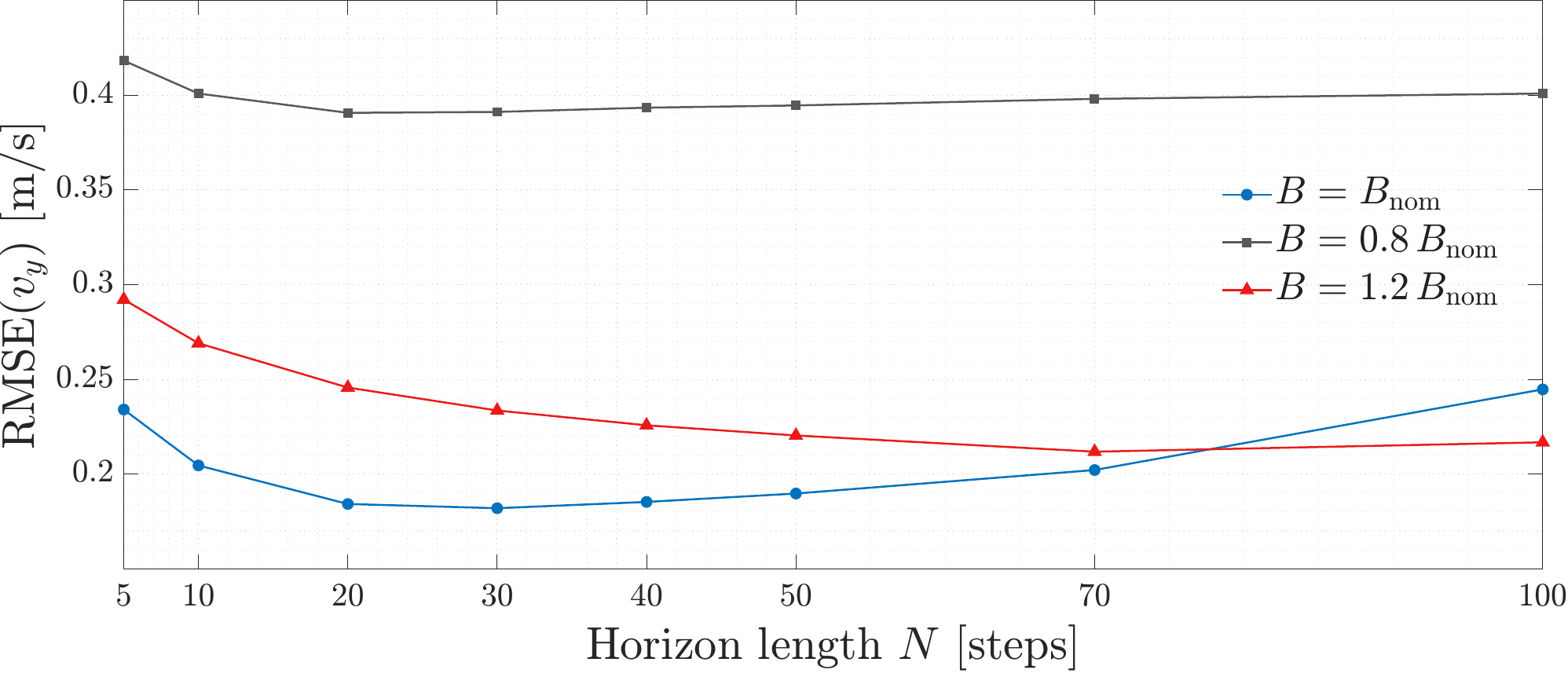}
    \caption{Sensitivity analysis of lateral-velocity RMSE with respect to horizon duration $T_h$ and SST curvature factor $B$.}
    \label{fig:rmse}
\end{figure}

On the same dataset, we evaluated the influence of the horizon length and of the SST curvature factor $B$ on the
estimation accuracy. The lateral-velocity RMSE is reported in Fig.~\ref{fig:rmse} as a function of the horizon duration
$T_h = N T_s$, with $T_s=0.01\,\mathrm{s}$ and $N\in\{5,10, 20,30,40,50,70,100\}$, corresponding to
$T_h\in\{0.05,0.10, 0.20,0.30,0.40,0.50,0.70,1.00\}\,\mathrm{s}$.

As anticipated from the SST discussion, excessively long horizons can degrade performance because the simplified sinusoidal tire approximation becomes less representative over wider slip-angle excursions, increasing accumulated model mismatch over the window considering that $\theta$ is constant over one horizon. Conversely, overly short horizons tend to increase RMSE because the problem becomes more sensitive to measurement noise and less constrained by the dynamic model.
The curvature factor $B$ is the main explicit shape parameter in the SST mapping. It is selected as discussed in Section~\ref{subsec:tire_model} to keep the sine mapping well-behaved over the slip-angle envelope observed in the data (cf. the monotonicity/validity range). In Fig.~\ref{fig:rmse}, the black and red curves correspond respectively to values 20\% below and 20\% above the nominal value.

Decreasing $B$ reduces the curvature of $\sin(B\alpha)$ over the operating slip range, making the surrogate less able to reproduce the observed nonlinear force build-up near the limit; increasing $B$ produces the opposite distortion (earlier curvature/saturation). In both cases, the RMSE increases compared to the nominal tuning, while the estimator remains stable and preserves the correct qualitative trends.

\section{CONCLUSION AND FUTURE WORK}
\label{section:conclusion}

This paper presented an IMU-centric Moving Horizon Estimation framework for vehicle lateral dynamics. By combining a constrained optimization-based formulation with a compact sine-based tire-force surrogate, the proposed approach reconstructs lateral velocity $v_y$ and lateral acceleration $a_y$, while providing consistent yaw-rate estimates from inertial measurements, measured inputs, and available longitudinal velocity. The framework operates without external odometry updates and without requiring high-dimensional tire-parameter identification. Experimental results show reduced sensitivity to tire-parameter uncertainty and changing grip compared with the evaluated filtering baselines, primarily because the estimation problem accommodates effective force-capacity variations through bounded and regularized variables rather than assuming a fixed tire model.

A practical advantage of the approach is its limited dependence on vehicle-specific tuning. In its reduced form, the formulation relies mainly on geometric parameters and mass, which are typically available or can be measured with limited effort. This is supported by results on the public \emph{REVS Program Vehicle Dynamics Database}, involving two human-driven sports cars, and on autonomous Super Formula runs with the \emph{EAV-25} platform. Compared with learning-based approaches, which may require large and condition-diverse training datasets and retraining when the vehicle, tires, or operating conditions change, the proposed method offers a physics-based alternative with limited per-vehicle calibration.

Measured solver runtimes on the tested high-end CPU platform suggest that real-time implementation is feasible.

Future work will address online adaptation of the SST curvature parameter $B$ to further reduce dependence on preset tire-model choices. Another direction is real-time embedded implementation, including the interaction between fast state estimation and slower force-capacity or grip adaptation. A promising architecture is a hierarchical estimator in which a high-rate lightweight filter provides fast state updates, while a lower-rate constrained optimization layer refines slowly varying parameters under richer models. We also plan to extend the validation of the proposed framework to a wider range of vehicles and operating scenarios, including urban and off-road environments.
\bibliographystyle{IEEEtran_DOI}
\bibliography{IEEEabrv} 

\end{document}